\documentclass[12pt]{article}

\newcommand{\lsim}{\mathrel{\mathop{\kern 0pt \rlap
  {\raise.2ex\hbox{$<$}}}
  \lower.9ex\hbox{\kern-.190em $\sim$}}}
\newcommand{\gsim}{\mathrel{\mathop{\kern 0pt \rlap
  {\raise.2ex\hbox{$>$}}}
  \lower.9ex\hbox{\kern-.190em $\sim$}}}
\newcommand{\gagamma}{g_{a\gamma\gamma}}

\input{psfig}

\begin{document}
\title{Particle Dark Matter and Solar Axion Searches\\ with a
small germanium detector at the \\Canfranc Underground Laboratory}

\date{}
\author{}
\maketitle

\vspace{-2cm}
\begin{center}

A.~Morales$^{a}$\footnote{corresponding author:
amorales@posta.unizar.es},
F.T.~Avignone~III$^{b}$, R.L.~Brodzinski$^{c}$, S.~Cebri\'{a}n$^{a}$,
\\ E.~Garc\'{\i}a$^{a}$, D.~Gonz\'{a}lez$^{a}$,
I.G.~Irastorza$^{a}$, H.S.~Miley$^{c}$, \\ J.~Morales$^{a}$,
A.~Ortiz~de~Sol\'{o}rzano$^{a}$, J.~Puimed\'{o}n$^{a}$,
\\J.H.~Reeves$^{c}$, M.L.~Sarsa$^{a}$, S.~Scopel$^{a}$,
J.A.~Villar$^{a}$\\

\vspace{0.2cm} {\it (The COSME Collaboration)}

\end{center}

\begin{center}
\begin{em}
\footnotesize

$^{a}$Laboratory of Nuclear and High Energy Physics, University of
Zaragoza, 50009 Zaragoza, Spain
\\
$^{b}$University of South Carolina, Columbia, South Carolina 29208
USA
\\
$^{c}$Pacific Northwest National Laboratory, Richland, Washington
99352 USA
\\

\end{em}
\end{center}

\begin{abstract}
A small, natural abundance, germanium detector (COSME) has been
operating recently at the Canfranc Underground Laboratory (Spanish
Pyrenees) in improved conditions of shielding and overburden with
respect to a previous operation of the same detector
\cite{Morales:1992,COSME1}. An exposure of 72.7 kg day in these
conditions has at present a background improvement of about one
order of magnitude compared to the former operation of the
detector. These new data have been applied to a direct search for
WIMPs and solar axions. New WIMP exclusion plots improving the
current bounds for low masses are reported. The paper also
presents a limit on the axion-photon coupling obtained from the
analysis of the data looking for a Primakoff axion-to-photon
conversion and Bragg scattering inside the crystal. \vspace{0.5
cm}


\end{abstract}

\section{Introduction}

Substantial evidence and well-founded arguments exist pointing out
that the Universe may well consist of a suitable mixture of cold
dark matter (CDM), hot dark matter (HDM) and baryons (in the
amount required by the primordial nucleosynthesis), which together
with a large component of dark energy could complete the proper
gravitational balance of the Universe. Regarding the nature of the
dark matter, there are compelling reasons to believe it consist
mainly of cold non-baryonic particles. Among these candidates,
Weakly Interacting Massive Particles (WIMPs) and axions are the
front runners. The lightest stable particles of supersymmetric
theories, like the neutralino, describe a particular class of
WIMPs\cite{Gri}. On the other hand, axions \cite{Peccei-Quinn},
the light pseudoscalar bosons emerging from the spontaneous
breaking of the Peccei-Quinn symmetry proposed to solve the strong
CP problem, are also favorite candidates to provide the dark
matter density for masses of the order of microelectronvolts
\cite{Raffelt:1990}.

The direct detection of WIMPs relies on the measurement of their
elastic scattering off the target nuclei of a suitable
detector\cite{Mor99}. The non relativistic and heavy (GeV -- TeV)
WIMPs supposedly filling (at least partially) the galactic haloes
could hit a Ge nucleus producing a nuclear recoil of a few keV.
Because of the small WIMP-matter interaction cross sections the
rate is extremely low ranging from 1 to 10$^{-2}$ c/kg/day
according to the type of interaction and type of WIMP. The
detection of such small and rare signals requires ultra--low
background detectors with very low energy thresholds. Many
experiments are underway using a variety of techniques to achieve
both goals. Germanium ionization detectors have reached one of the
lowest raw background levels of any type of detector and have a
reasonable ionization yield($\sim 0.25$). Should they have
sufficiently low energy thresholds, they would be attractive
devices for WIMP direct detection. This paper presents the case of
one of such germanium detectors of rather low energy threshold and
fairly low background rate.

As it is well known, solar axions can been searched for with
crystal detectors through the axion-to-X ray Primakoff conversion
with coherent Bragg diffraction
\cite{Paschos:1994yf,Creswick:1998}. This paper also deals with a
search for such axions with the same germanium detector, COSME.


\section{Experimental setup}
The COSME detector is a p-type coaxial hyperpure natural germanium
crystal, of dimensions 22 mm (length) $\times $ 52.5 mm (diameter)
corresponding to an active volume of 44 cm$^3$ and a mass of 234 g
which has a long-term resolution of 0.43 keV full width at half
maximum (FWHM) at 10.37 keV. The detector was specially built for
dark matter searches, i. e., with a low energy threshold and
ultralow background materials. The cryostat, of 1.5 mm thick
electroformed copper, was made by Pacific National Northwest
Laboratory (PNNL). The field effect transistor (FET) is shielded
with 450 year old (Spanish) lead.

The COSME detector was formerly used in the Canfranc tunnel in one
of the first dedicated searches for WIMPs \cite{Morales:1992,
COSME1}. Now it has been placed in a deeper location in the same
tunnel within a significantly improved shielding, the same as the
IGEX experiment \cite{IGEX2b, IGEXDM}. The innermost shield of the
IGEX and COSME detectors consists of 2.5 tons of archaeological
lead 2000-year-old forming a 60 cm cube. The detector fits into a
precision-machined hole in this lead block to minimize the empty
space around the detector available to radon. Nitrogen gas, at a
rate of 140 liters per hour, evaporating from liquid nitrogen, is
forced into the detector chamber to create a positive pressure and
to further avoid radon intrusion. The archaeological lead block is
located inside of another shield made of bricks of
low-activity-lead ($\sim$ 10 tons) forming all together a cube of
1 m on a side. A 2-mm-thick cadmium sheet surrounds the external
lead shield, and two layers of plastic seal the assembly tightly.
A cosmic ray muon veto covers the top and two sides of the
shielding. Finally, an external polyethylene neutron moderator 20
cm thick (1.5 tons) covering the six sides of the cube complete
the shield. The entire set-up is supported by an iron structure
resting on noise-isolation blocks. The experiment is located in a
room isolated from the rest of the laboratory which has an
overburden of 2450 m.w.e., reducing the muon flux to a (measured)
value of $2\times 10^{-7} $cm$^{-2}$s$^{-1}$.

Further details about the detector are given in Ref.\cite{COSME1}
whereas the setup and shielding are described in the IGEX papers
\cite{IGEX2b,IGEXDM}. The data acquisition and filtering
techniques are similar to the previous experiment
\cite{Morales:1992,COSME1} but with the addition of muon vetoes
working in anticoincidence as in IGEX \cite{IGEXDM}.

\section{Experimental Results}
\begin{figure}[tb]
\psfig{figure=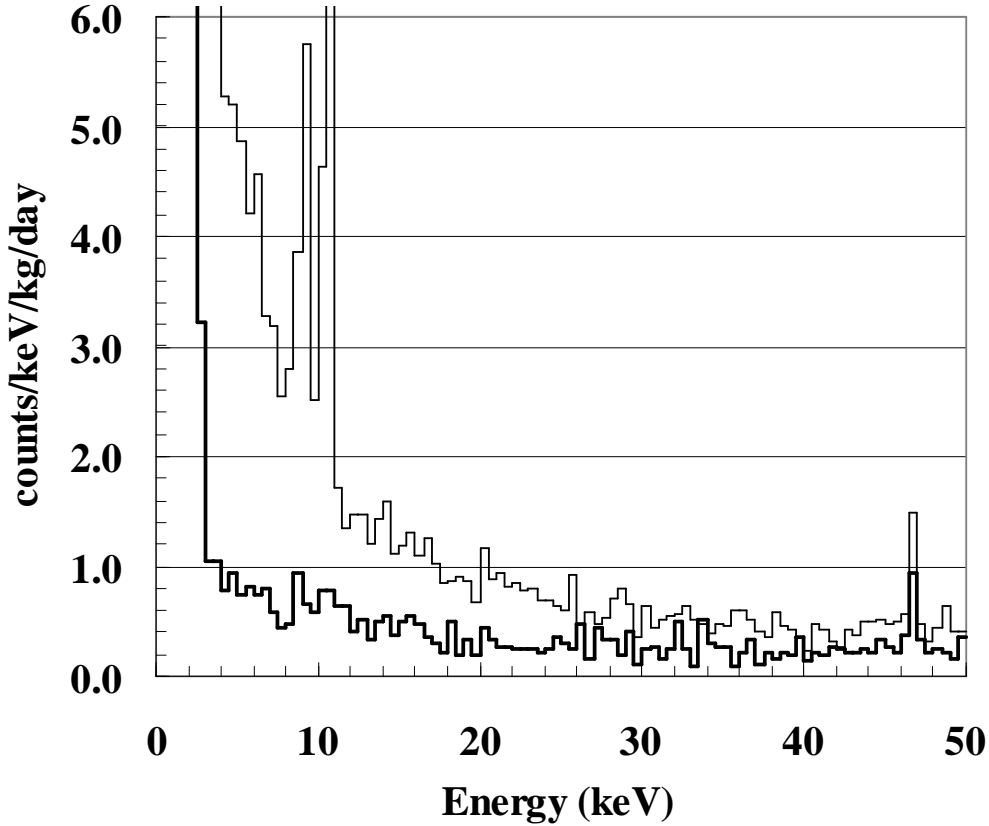,width=140mm}

\caption{\small The thick line shows the data reported in this
paper, obtained with 72.7 kg days of exposure with COSME detector.
The thin line shows the previous running of the same detector
\cite{COSME1} for comparison.}

\label{espectro}

\end{figure}
\subsection{Spectrum}

The COSME detector has been running during the years 1998 and 1999
in these new conditions, and has accumulated data corresponding to
a statistics of 72.7 kg days of exposure. Figure \ref{espectro}
gives the experimental differential rate corresponding to this
acquisition (thick line) compared to the previous set of data from
1990 to 1992 \cite{COSME1} (thin line). A significant decrease of
the background with respect to the former data is clearly shown.
It is also worth noticing the near disappearance of the peaks at
8.98 keV and 10.37 keV corresponding to cosmogenically induced
$^{65}$Zn and $^{68}$Ge isotopes (T$_{1/2}$=244.3 d and 288 d EC
X-rays from $^{65}$Cu and $^{68}$Ga respectively) after the
long-term underground storage of the detector. Also the 46.5 keV
peak of $^{210}$Pb shows a significant decrease.

The energy threshold achieved in this new setting is 2.5 keV and
the long-term energy resolution obtained in the visible low energy
peaks (Pb and Bi X-rays and 46.5 keV from $^{210}$Pb) was the same
as in the previous run. The background in the low energy region
(2.5 - 10 keV) is 0.7 c/keV/kg/day, an improvement of almost one
order of magnitude compared with the previous run. The background
goes down to 0.4 c/keV/kg/day in the 12--20 keV region, and 0.3
c/keV/kg/day in the 20--30 keV region. Although this background is
not as low as that obtained in the other detectors sharing the
same shielding (IGEX\cite{IGEXDM}), its better threshold leads to
an improved bound for low mass WIMP exclusion.

\subsection{WIMP exclusion plot}

As previously stated, the new data have been analyzed first to get
a limit on the WIMP-nucleon cross section as a function of the
WIMP mass. The exclusion plot is derived from the recorded
spectrum in 0.5-keV bins (from the 2.5 keV threshold up to 50
keV), by requiring the theoretically predicted signal in an energy
bin to be less than or equal to the (90\% C.L.) upper limit of the
(Poisson) recorded counts. The derivation of such an interaction
rate signal assumes standard hypothesis and parameters, i.e., that
the WIMPs form an isotropic, isothermal, non-rotating halo of
density $\rho=0.3 $ GeV/cm$^3$, which has a maxwellian velocity
distribution with $v_{rms}=270$ km/s (with an upper cut
corresponding to an escape velocity of 650 km/s), and a relative
Earth-halo velocity of $v_r=230$ km/s). Other, more elaborated
halo models, as well as other kinematical parameters would lead to
different results. The cross sections are normalized per nucleon
assuming a dominant scalar interaction, as the most likely
observable coupling with the current experimental sensitivities:

\begin{equation}\label{norm}
    \sigma_{N\chi} = \sigma_{n\chi}A^2 \frac{\mu^2_{W,N}}{\mu^2_{W,n}}
\end{equation}

\noindent where $A$ is the target (germanium) mass number,
$\mu^2_{W,N}$ is the WIMP-nucleus reduced mass, and $\mu^2_{W,n}$
the WIMP-nucleon reduced mass. The Helm parameterization
\cite{Helm} is used for the scalar nucleon form factor. The
quenching factor has been assumed to be equal to 0.25, which is a
good approximation if the relevant bins for the calculation of the
exclusion plot are below $\sim 10$ keV, as is the case of COSME.

\begin{figure}[tb]
\psfig{figure=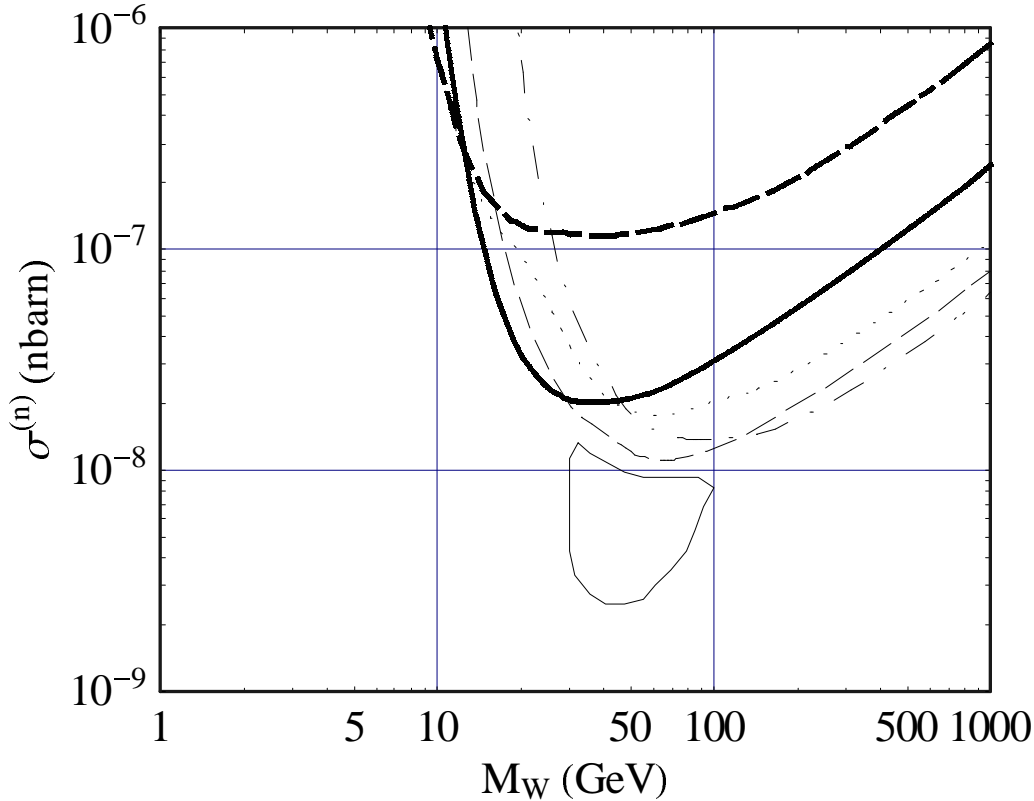,width=140mm}
\caption{\small Exclusion plot for spin-independent interaction
where bounds coming from different experiments are drawn. The
thick line shows the exclusion obtained with the data reported in
this paper, obtained with 72.7 kg-days of exposure with COSME
detector.  The thick dashed line shows the one from the previous
running of the same detector \protect\cite{COSME1} for comparison.
Limits from a recent result of IGEX-DM
experiment\protect\cite{IGEXDM} (thin-dashed), from the
Heidelberg-Moscow experiment\protect\cite{Bau} (dot-dashed) and
from a combination of previous germanium
experiments\cite{COSME1,Gotthard,TWIN} (dots) are also shown. The
closed region corresponds to the (3$\sigma$) annual modulation
effect reported by the DAMA collaboration (including NaI-1,2,3,4
runnings)\protect\cite{Ber99}.}

\label{exclusion}

\end{figure}

The exclusion plot derived from the data reported here following
such a procedure is shown in figure \ref{exclusion} by a thick
solid line. This result is compared with the limit obtained with
the previous data of COSME (thick dashed line) as well as a
combined contour obtained from several previous germanium
experiments \cite{COSME1,Gotthard,TWIN} (thin dotted line) and two
recent germanium results, IGEX-DM \cite{IGEXDM} (thin dashed line)
and Heidelberg-Moscow \cite{Bau} (thin dot-dashed line). All the
curves have been computed from the raw data with the same set of
assumptions (except for those for which a 0.25 quenching factor
was not a good aproximation, for which an appropriate energy
dependence has been used \cite{Bau}). The DAMA region
corresponding to its reported annual modulation effect
\cite{Ber99} is also shown. As can be seen, the new COSME plot
improves other germanium bounds in the 12--30 GeV region and
together with the other Canfranc results from Ge detectors (COSME
I and IGEX-DM) provide the best bound below 200 GeV obtained. up
to now with conventional germanium ionization detectors. Better
exclusion contours have been obtained with NaI scintillators which
use statistical pulse shape background discrimination
\cite{Ber98}, as well as with germanium bolometers which
simultaneously detect ionization and allow an event-by-event
background rejection\cite{CDMS}.

\subsection{Limit on axion-photon coupling}

If axions exist they could be copiously produced in the core of
the stars from where they can extract energy efficiently competing
with other conventional mechanisms. The cooling of stars by
axionic energy emission could affect the star's properties and its
evolution and so, to avoid conflict with the experimental
observation, strong constrains on the coupling strength or mass of
the stellar axion follow \cite{Raffelt:1990}. The discovery of
axions, if they exist, --either galactic or stellar-- is one of
the challenges of current astrophysics and various efforts are
underway. As far as this second category of axions is concerned, a
nearby and powerful source of stellar axions would be the Sun.

Crystal detectors provide a simple mechanism for solar axion
detection \cite{Paschos:1994yf,Creswick:1998}. In fact, axions can
pass in the proximity of the atomic nuclei of the crystal where
the intense electric field can trigger their conversion into
photons. In the process the energy of the outgoing photon is equal
to that of the incoming axion. The axion production rate in the
Sun --through Primakoff conversion of the blackbody photons in the
solar plasma-- can be easily estimated
\cite{vanBibber:1989ge,Creswick:1998} within the standard solar
model, resulting in an axion flux of an average energy of about 4
keV that can produce detectable X-rays when reconverted again in a
crystal detector. Depending on the direction of the incoming axion
flux with respect to the planes of the crystal lattice, a coherent
effect can be produced when the Bragg condition is fulfilled, with
the ensuing strong enhancement of the signal.

A useful parametrization\cite{Creswick:1998} of the solar axion
flux at Earth is the following:
\begin{equation}
\frac{d\Phi}{dE_a}=\sqrt{\lambda}\frac{\Phi_0}{E_0}\frac{(E_a/E_0)^3}{e^{E_a/E_0}-1}
\label{eq:flux}
\end{equation}
\noindent where $\lambda$=($\gagamma\times 10^8/$${\rm
GeV}$$^{-1}$)$^4$ is a dimensionless coupling introduced for later
convenience, $\Phi_0$=5.95 $\times$$10^{14}$ cm$^{-2}$ sec$^{-1}$
and $E_0$=1.103 keV.

\begin{figure}[t]
\begin{center}
\mbox{\psfig{figure=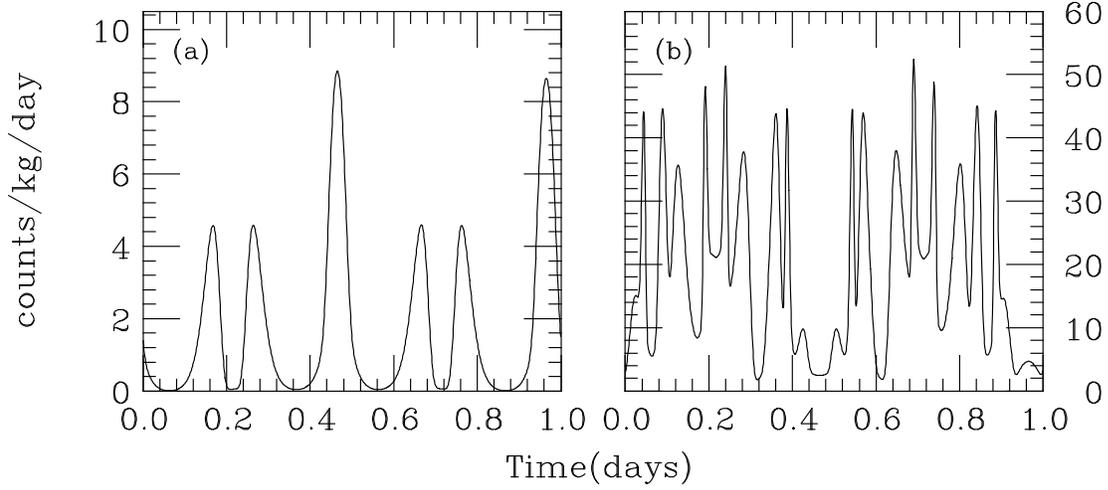,width=150mm}}
\caption{Expected axion signals for coherent Primakoff conversion
in a germanium crystal as a function of time for $\lambda=1$, for
two different energy windows: a) 2 keV$\leq E_{ee}\leq$2.5 keV; b)
4 keV$\leq E_{ee}\leq$4.5 keV. \label{fig:axions}}
\end{center}
\end{figure}

Making use of eq. \ref{eq:flux}, as well as the cross-section of
the process and appropriate crystallographic information, we have
calculated the expected axion-to-photon conversion count rate
$R(t)$ in a germanium detector (See ref.\cite{Cebrian:1999,
Creswick:1998} for further details). An example of the expected
count rate is shown in figure \ref{fig:axions} as a function of
time during one day. As expected, the signal shows a strong time
dependence throughout the day, due to the motion of the Sun in the
sky. This correlation of the expected rate with the position of
the Sun in the sky is a distinctive signature of the axion which
can be used, at the least, to improve the signal/background ratio.

To extract all this information from a experimental set of data we
introduce, following Ref. \cite{SOLAX}, the quantity:
\begin{equation}
\chi=\sum_{i=1}^{n} \left[ \bar{R}(t_i)-<\bar{R}>\right]\cdot
n_i\equiv \sum_{i=1}^{n} W_{i} \cdot n_{i} \label{eq:chi}
\end{equation}
\noindent where $\bar{R}(t) = R(t)/\lambda$, the $n_i$ indicates
the number of measured events in the time bin $t_i,t_i+\Delta t$
and the sum is over the total period $T$ of data taking. The
brackets indicate time average.

By definition the quantity $\chi$ is expected to be compatible
with zero in absence of a signal, while it weights positively the
events recorded in coincidence with the expected peaks.

The time distribution of $n_i$ is supposed to be Poissonian, with
mean:
\begin{equation}
<n_i>=\left[ \lambda \bar{R}(t_i)+b \right]\Delta t.
\label{eq:mediaconti}
\end{equation}
Assuming that the background $b$ dominates over the signal the
expected average and variance of $\chi$ are given by:
\begin{eqnarray}
<\chi>&=&\lambda \cdot A\\ \sigma^2(\chi)&\simeq& b/A
\label{eq:medie}
\end{eqnarray}
\noindent with $A\equiv \sum_i W_i^2\Delta t$. Each energy bin
$E_k,E_k+\Delta E$ with background $b_k$ provides an independent
estimate $\lambda_k=\chi_k/A_k$ so that one can get the most
probable combined value of $\lambda$:
\begin{eqnarray}
\lambda&=&\sum_k\chi_k/\sum_l A_l \nonumber \\
\sigma(\lambda)&=&\left(\sum_k A_k/b_k\right)^{-\frac{1}{2}}.
\label{eq:lambda}
\end{eqnarray}

\begin{figure}[t]
\begin{center}
\mbox{\psfig{figure=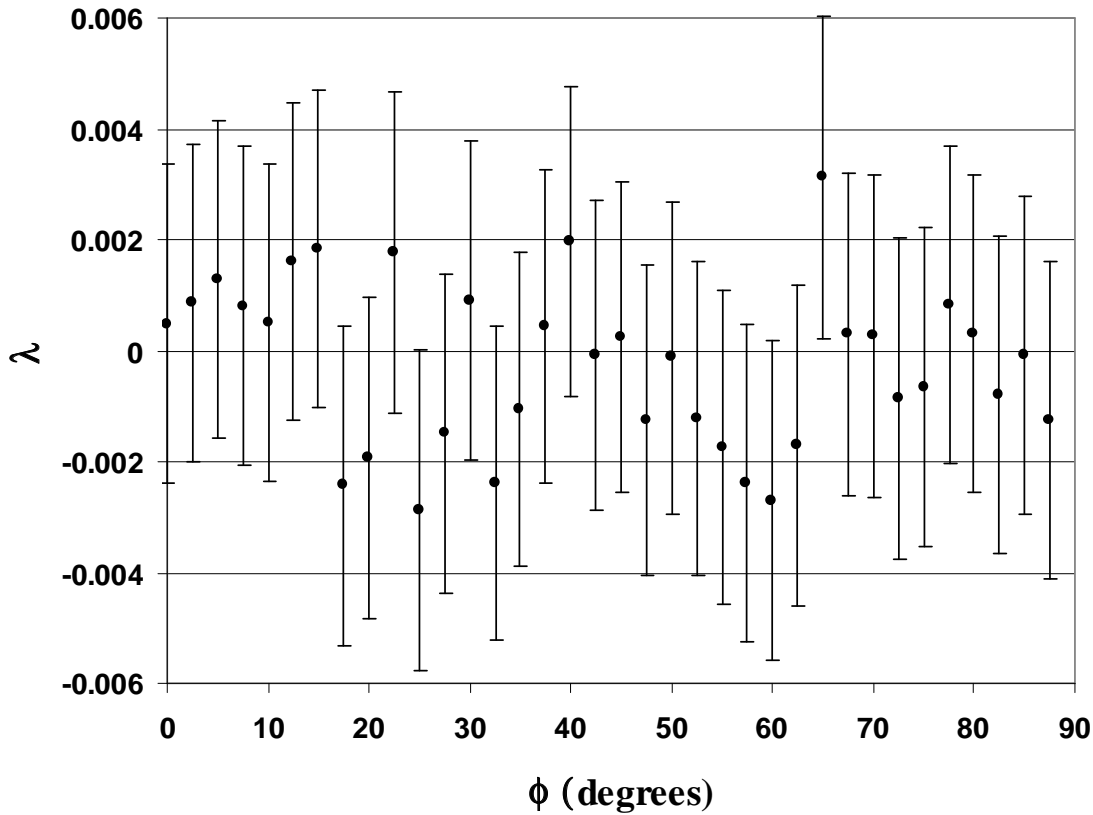,width=150mm}}
\caption{Values of $\lambda$ obtained for different assumed
rotations around the fixed vertical axis.\label{fig:lambdas}}
\end{center}
\end{figure}

This method provides an estimation of the axion-photon coupling
$\gagamma$ if an axion signal is present in the experimental data
(and is accessible to the sensitivity of the experiment), or an
upper limit to $\gagamma$ in case of a negative result. If the
exact orientation of the crystal is not known we must repeat the
calculation for all possible ones, and take the most conservative
result. We know only the direction of one of the crystallographic
axis of COSME, which is vertical with respect to the laboratory
frame, so we have repeated the calculation for all possible
rotations around the fixed vertical axis. The different values for
$\lambda$ so obtained are shown in figure \ref{fig:lambdas} with
the corresponding 90\% error derived from $\sigma(\lambda)$. All
of them are compatible with absence of signal, so taking the most
conservative value, we obtain an upper limit $\lambda < 0.006$ at
95\% C.L., which corresponds to a limit on $\gagamma$:

\begin{equation}\label{limit}
  \gagamma < 2.78 \times 10^{-9} {\rm GeV}^{-1}
\end{equation}

The application of the statistical analysis described above can be
viewed as a background rejection technique, which in our case
results in a reduction of about two orders of magnitude. The limit
presented here is practically equal to that obtained by the SOLAX
Collaboration with another Ge detector\cite{SOLAX}. These limits
are the \emph{mass independent} (although solar model dependent)
most stringent laboratory bounds for the axion-photon coupling
obtained so far. A recent result, however, obtained with the Tokyo
helioscope \cite{tokyo} (a $\sim$ 2 m long superconducting magnet
which converts axions into photons in a $\sim$ 4 Tesla magnetic
field), improves by a factor 3-5 the COSME and SOLAX bounds for
axion masses below 0.26 eV.

\section{Conclusion}
A small (234 g) germanium detector of natural isotopic abundance,
with an energy threshold of 2.5 keV and an energy resolution of
430 eV at the 10.37 keV X-ray $^{68}$Ga peak, has collected data
up to a statistics of 311 days, in a search for WIMPs. The timing
information of the same set of data has allowed also a search for
the characteristic time pattern of an hypothetical solar axion
signal as expected from the relative motion of the detector with
respect to the Sun along the day. Both searches have provided
bounds for the couplings $\sigma_W(m)$ and $\gagamma$ of both
types of hypothetical particles to matter and radiation
respectively. The obtained results have been proven to be
competitive (or superior) with other current searches of such
types of particles, which use similar techniques.

\section*{Acknowledgements}
The Canfranc Astroparticle Underground Laboratory is operated by
the University of Zaragoza under contract No. AEN99-1033. This
research was partially funded by the Spanish Commission for
Science and Technology (CICYT), the U.S. National Science
Foundation, and the U.S. Department of Energy.

\end{document}